\documentclass[prb,showpacs,twocolumn,superscriptaddress,floatfix]{revtex4}
\usepackage{amsmath}
\usepackage{graphicx}
\usepackage{float}
\usepackage{amssymb}
\usepackage{times}
\DeclareGraphicsExtensions{.eps,.ps}

\def\bi         {\begin{itemize}}
\def\ei         {\end{itemize}}
\def\benu	{\begin{enumerate}}
\def\eenu	{\end{enumerate}}
\def\bmat       {\left[ \begin{array}}
\def\emat       {\end{array} \right]}
\def\beq	{\begin{equation}}
\def\eeq	{\end{equation}}
\def\beqn       {\begin{eqnarray*}}
\def\eeqn       {\end{eqnarray*}}
\def\beqa       {\begin{eqnarray}}
\def\eeqa       {\end{eqnarray}}
\def\bquote	{\begin{quote}}
\def\equote	{\end{quote}}

\def\bwide	{\begin{widetext}}
\def\ewide	{\end{widetext}}

\begin{document}

\title{Supersolidity from defect-condensation in the extended boson Hubbard model}

\author{Yu-Chun Chen}
\affiliation{Department of Physics, National Taiwan University, Taipei, Taiwan 106}

\author{Roger G. Melko}
\affiliation{Department of Physics and Astronomy, University of Waterloo, Ontario, N2L 3G1, Canada}
\affiliation{Materials Science and Technology Division, Oak Ridge National Laboratory,
Oak Ridge TN, 37831}
\author{Stefan Wessel}
\affiliation{Intitut f\"{u}r Theoretische Physik III, Universit\"{a}t  Stuttgart, 70550 Stuttgart, Germany}
\author{Ying-Jer Kao}
\email{yjkao@phys.ntu.edu.tw}
\affiliation{Department of Physics, National Taiwan University, Taipei, Taiwan 106}
\affiliation{Center for Theoretical Sciences, National Taiwan University, Taipei, Taiwan 106}
\date{\today}

\begin{abstract}
We study the ground state phase diagram of the hard-core extended boson Hubbard 
model on the square lattice with both nearest- (nn) and
next-nearest-neighbor (nnn)  hopping and repulsion, using Gutzwiller mean field theory and quantum Monte Carlo 
simulations.
We observe the formation of supersolid states with checkerboard, striped, and quarter-filled crystal structures, 
when the system is doped away from 
commensurate fillings.  
In the striped supersolid phase, a strong anisotropy in the superfluid density is obtained from the simulations;
however, the transverse component remains finite, indicating a true two-dimensional superflow. 
We find that upon doping, the striped supersolid transitions directly 
into the supersolid with quarter-filled crystal structure, via a first-order stripe melting transition.

\end{abstract}

\maketitle


\section{Introduction} 
In 1956, Penrose and Onsager~\cite{Penrose56PR} first posed the question
of whether one could expect superfluidity in a solid --
a supersolid state -- with coexisting diagonal 
and off-diagonal long-range order.  
They showed that for a perfect crystal, where the 
wave function of the particles is localized near each lattice site, superfluidity
does not occur at low temperature. Later it was proposed~\cite{Andreev69JETP,Chester70PRA,Leggett70PRL} that fluctuating  
defects in {\em imperfect} crystals can condense to form a superfluid, and a supersolid state 
(with both superflow and periodic modulation in the density) emerges. 
In 2004, Kim and Chan reported signatures of superfluidity in solid $^4$He in torsional oscillator experiments, \cite{Kim04Nature}
where a drop in the resonant period, observed at around $T \sim 0.2$K, suggested the existence of a non-classical rotational inertia in the crystal.\cite{Leggett70PRL}
Following the discovery by Kim and Chan, many experiments and theories have attempted to explain this fascinating observation;  the situation remains, however, controversial.
\cite{Ceperley04PRL,Prokofev05PRL,Burovski05PRL,Kim06PRL,Rittner06arxiv,day:105304}

On the other hand, with improvements of quantum Monte Carlo (QMC) methods, the origin of 
supersolid phases can be studied exactly in both continuum and lattice models.  
Exotic quantum phases, including supersolids, are highly sought-after in lattice models,
particularly those that may be realized by 
loading ultra-cold bosonic atoms onto optical lattices.\cite{Anderson95Science,Greiner02Nature}  The generation of a Bose-Einstein condensate (BEC) 
in a gas of dipolar 
atoms \cite{Griesmaier05PRL} with longer-range interactions provides one promising route to search for the supersolid state.\cite{dipolar_theory}
The extended boson Hubbard model is the obvious microscopic Hamiltonian to study these
systems, and
supersolids have been found in this model on various lattices.
\cite{Batrouni95PRL,Hebert01PRB,Sengupta05PRL,triangular_SS,kagome_SS}
A simplified phenomenological picture for understanding the supersolid phase in these models 
has been the aforementioned ``defect-condensation'' scenario: starting from a perfect lattice crystal
at commensurate filling, supersolidity arises when dopants (particles or holes)
condense and contribute a superflow.  In the simplest scenario -- hard-core bosons doped above
commensurate filling -- this phenomenological picture suggests 
``microscopic phase separation'' between the crystal and superfluid sublattices.
Recent work on triangular lattice supersolids \cite{Melko06PRB,Tizzle} has called this 
simple interpretation into question, since there one apparently finds examples
where particles on the crystal lattice also take part in the superflow. However, 
frustration complicates
the interpretation of these results, since the underlying order-by-disorder mechanism 
facilitates supersolid formation at half-filling.

To more closely study the degree to which defect-condensation
plays a role in the mechanism behind lattice supersolids,  we study the extended 
hard-core boson Hubbard model with both nearest-neighbor (nn) and next-nearest-neighbor (nnn) hopping
and repulsive interactions on the square lattice. 
The Hamiltonian is
\begin{eqnarray}
\mathcal{H}&=&-t\sum_{\langle i,j\rangle }(a_i^\dagger a_j+a_ia_j^\dagger)+V_1
\sum_{\langle i,j\rangle } n_in_j -\mu\sum_i n_i \nonumber\\
& &-t'\sum_{\langle\langle i,j\rangle\rangle }(a_i^\dagger a_j+a_ia_j^\dagger )+V_2\sum_{\langle\langle i,j\rangle\rangle } 
n_in_j, \label{Ham}
\end{eqnarray}
where $a_i^\dagger$ and $a_i$ are the boson creation and annihilation operators, $n_i=a_i^\dagger a_i$ is the number operator, and
$\langle i,j \rangle$ denotes the nearest- and 
$\langle\langle i,j\rangle\rangle$  next-nearest neighbors.     
The $t'=0$ limit has been studied previously, and is known to harbor several crystal solids, 
including a checkerboard structure that upon doping is unstable towards phase separation.\cite{Batrouni00PRL, Hebert01PRB}
A striped crystal and a stable striped supersolid were also found in the $t'=0$ limit.\cite{Hebert01PRB}

Here, we study the full Hamiltonian of Eq.~(\ref{Ham}) using both Gutzwiller mean field theory
(Section~\ref{sGMF}) and stochastic series expansion (SSE)
quantum Monte Carlo (QMC) simulations (Section~\ref{sQMC}) based on the directed loop algorithm.\cite{Sandvik91PRB,Syljuasen02PRE}
We confirm that the model contains a variety of lattice crystals, including a checkerboard and 
striped phase at half-filling, plus a \textit{quarter-filled} solid.\cite{Schmid04phd}  The nnn-hopping $t'$ is found 
to stabilize a checkerboard supersolid away from half-filling; this and other supersolid phases
with differently broken symmetry are studied in detail. In general, we find that although 
supersolid phases are not stabilized at commensurate fillings, they are readily formed  upon doping.
However, we demonstrate that, contrary to the simple phenomenological picture of 
defect-condensation, where the crystal and superfluid sublattices are clearly distinct,
in at least one of our supersolid phases particles from the crystal sublattice also 
participate to a large degree in the superflow.

\section{Gutzwiller mean field approximation}
\label{sGMF}

We begin by surveying the ground state phase diagram of the model in several limits using
Gutzwiller mean-field theory.  The
Gutzwiller variational method\cite{Gutzwiller63PRL} is a powerful technique 
for studying strongly correlated system. The ground state of an interacting 
system is constructed from the corresponding noninteracting ground state, 
\begin{equation}
|\phi_g\rangle = \prod_{i}\left( \sum_{n_i}f_{n_i} |n_i \rangle\right).
\end{equation}
Here, the $f_{n_i}$ are  site-dependent 
variational parameters, which can be optimized 
via minimizing the energy
\[
E_0=\frac{\langle \phi_g | H |\phi_g\rangle}{
\langle \phi_g |\phi_g\rangle}
\]
 of the 
variational state. The $|n_i\rangle$ form the local Fock basis at site $i$ with 
$n_i$ particles in state $|n_i\rangle$ . In the hardcore limit, 
we only need to keep the local states $|0\rangle$ 
and $|1\rangle$. Physical quantities are calculated within the 
variational ground states. In particular, we measure the local density $\langle n_i \rangle $,
the density structure factor at wave vector $\mathbf{q}$
\begin{equation}
S({\bf q}) = \frac{1}{N} \sum_{i,j}
{\rm e}^{i\,{\bf q}\cdot({\bf r}_i - {\bf r}_j)}
\langle n_i n_j \rangle,
\label{Szstr}
\end{equation}
and superfluidity due to a finite value of $\langle a_i \rangle$.
The coexistence of superfluid order and Bragg peaks in the structure factor signifies supersolidity.

\begin{figure}[tb]
\begin{center}
\includegraphics[angle=0,width=3.4in,clip]{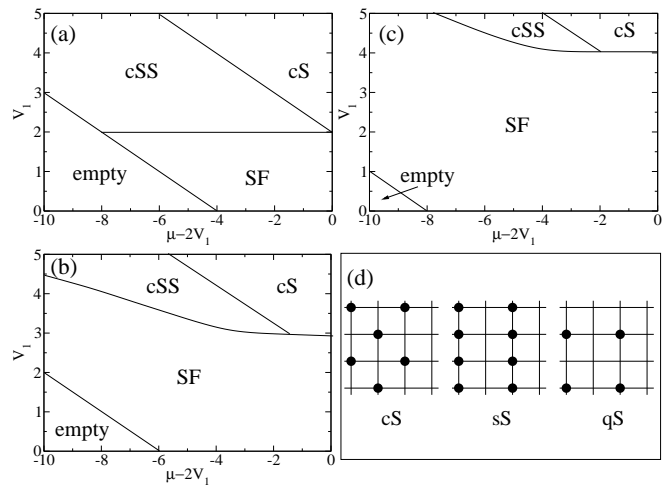}
\end{center}
\caption{
Mean-field phase diagram of $t-t'-V_1$ model at (a) $t=0$, (b) $t=0.5$, and
(c) $t=1.0$. See text for discussion of the phases. (d) possible quantum solid configurations: checkerboard
solid (cS), striped solid (sS) and quarter-filled solid (qS). All quantities are in units of $t'$.
}
\label{fig:MF1}
\end{figure}

We begin by studying the phase diagram within the Gutzwiller approximation 
near half-filling in the absence of the nnn 
repulsion
(i.e. for $V_2=0$).  From Fig.~\ref{fig:MF1}, we find that at the mean-field level, various phases
are stabilized already 
by this restricted set of parameters. These include a uniform superfluid (SF), a checkerboard solid (cS) with 
ordering
wave vector ${\bf q}=(\pi,\pi)$ (see Fig.~\ref{fig:MF1}d for an illustration), 
and in particular a checkerboard supersolid (cSS) - with coexisting diagonal order and
superfluidity, away from half-filling ($\mu=2V_1$). 

As expected, increasing the nn hopping $t$ destroys the solidity of the system. 
In particular, the supersolid region, found
when the system is doped away from half-filling, becomes a uniform
superfluid at large $t$. 
This clearly indicates, that large $t$ destabilizes the supersolid state. 

\begin{figure}[tb]
\begin{center}
\includegraphics[angle=0,width=3.4in,clip]{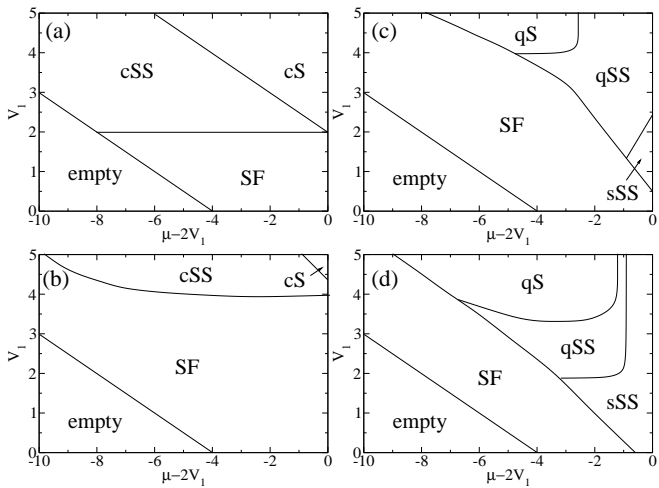}
\end{center}
\caption{
Mean-field phase diagram of $t-t'-V_1-V_2$ model at (a) $x=0: t=0,t'=1.0,$ and
$V_2=0$, (b) $x=0.3: t=0.3,t'=0.7$ and $V_2=1.5$, (c)
$x=0.7: t=0.7,t'=0.3$ and $V_2=3.5$ and (d) $x=1.0:t=1.0,t'=0$ and $V_2=5$. All quantities are in units of $t+t'$.
}
\label{fig:MF2}
\end{figure}

To study the effects of a finite nnn repulsion $V_2$, we first identify two limiting cases. For the $t-V_2$ model, a stable supersolid state is the 
striped supersolid (sSS)
\cite{Batrouni95PRL,Batrouni00PRL,Sengupta05PRL,SchmidPRL04} with ordering wavevectors
${\bf q}=(\pi,0)$ or $(0,\pi)$ if obtained.  
For the $t'-V_1$ model, a stable supersolid state is the checkerboard 
supersolid.  
To capture the behavior of the system between these limiting regimes, we introduce 
a parameter $0 \leq x \leq 1$, which interpolates between the two regions, by setting $ t=x, t'= 1-x,$ and $V_2 = 5x$ 
(see Ref.~[\onlinecite{Schmid04phd}]). 
In the following, we thus work in units of $t+t'=1$. 
Figure~\ref{fig:MF2} shows Gutzwiller mean field phase diagrams for different values of $x=0, 0.5$, and 1. 
While for finite $V_2$, half-filling is obtained for $\mu=2V_1+2V_2$, 
we still take $\mu-2V_1$ as the abscissa, in order to ease a direct comparison to the previous case of $V_2=0$.

The coexistence of both nn and nnn repulsions is expected to stabilize various solid states \cite{Schmid04phd}: 
at half-filling, with both $V_1$ and $V_2$ large, a checkerboard (striped) 
solid is formed for $V_1 > 2 V_2 (V_1 < 2 V_2) $\cite{Batrouni95PRL}; we find that 
for $\rho=1/4$, a quarter-filled solid (qS) (shown in Fig.~\ref{fig:MF1}d) emerges. 
In order to distinguish the different solids, we measure the structure factors
at reciprocal lattice vectors $\mathbf{q}=(\pi,\pi)$, $(\pi,0)$ and
$(0,\pi)$. 
 In addition, for the striped structure where a strong anisotropy due to the broken rotational 
symmetry occurs, the magnitude of the
difference between $O_x=S(\pi,0)$ and $O_y=S(0,\pi)$ is almost equal to the sum $O_x+O_y$. 
Keeping track of this quantity allows us to easily distinguish supersolids with an underlying striped
crystal, from supersolids with an underlying quarter-filled crystal, in which case all three structure factors become finite, but the 
difference $|O_x-O_y|$ is zero (see Table~\ref{order} for a summary).  

\begin{table}[tb]
\begin{center}
\begin{tabular}{|c||c|c|c|}
\hline
 Structure factor & checkerboard  & striped          & quarter-filled\\
\hline
   $S(\pi,\pi)$   &   $\neq 0$      & 0                & $\neq$ 0\\
   $O_x+O_y$      &    0          & $\neq$ 0           & $\neq$ 0\\
   $|O_x-O_y|$    &    0          & $\simeq O_x+O_y$ & 0\\
\hline
\end{tabular}
\end{center}
\caption{Order parameters associated with the different crystal orders.}
\label{order}
\end{table}

The case $x=0$, shown in Fig.~\ref{fig:MF2}a  corresponds to the $t'-V_1$ model, which we already discussed (compare to  
Fig. 1a). 
At $x=0.3$ (Fig.~\ref{fig:MF2}b), the superfluid phase expands, as the 
introduction of $t$ and $V_2$  destabilizes the checkerboard solid. 
For $x=0.7$ (Fig.~\ref{fig:MF2}c), with the model parameters approaching the $t-V_2$ limit,
the checkerboard structure disappears. Instead, striped and quarter-filled structurs emerge, including a 
striped supersolid (sSS), and a
quarter-filled supersolid (qSS).\cite{qSS} 
In the limiting case $x=1$ (Fig.~\ref{fig:MF2}d), we find two
different transition paths from the sSS to the superfluid upon doing. When $V_1<2$, the striped 
supersolid enters the superfluid directly. When $V_1>2$, the qS solid 
and qSS supersolid regions are passed as an intermediate regime, separating the striped supersolid sSS and the 
superfluid.

Clearly, these mean-field phase diagrams provide evidence for not only the existence of various different supersolid 
phases, 
but also for the possibility of direct quantum phase transitions between them (in particular between the qSS and sSS 
phases).  
In the next section, using these mean-field results as guidance, we turn to quantum
Monte Carlo simulations in order to 
study in details the various supersolid phases, as well as the transitions between them.

\section{Quantum Monte Carlo results}
\label{sQMC}

We performed extensive quantum Monte Carlo (QMC) simulations of the Hamiltonian Eq.~(\ref{Ham}) 
using a variation of the stochastic series expansion framework with directed loops.\cite{Sandvik91PRB,Syljuasen02PRE}  Correlation functions of density 
operators are easily measured within the QMC, and crystal order is signified by peaks in the ${\bf q}$-dependent 
structure factor of Eq.~(\ref{Szstr}).  The superfluid density is measured in the standard way in terms of winding 
number fluctuations, 
\begin{equation}
\rho_s^{{\bf a}} = \frac{\langle W_{{\bf a}}^2 \rangle}{\beta},
\label{RHOs}
\end{equation}
where ${\bf a}$ labels the $x$ or $y$ direction and $\beta$ is the inverse temperature.  Typically the stiffness is averaged over both directions,
unless measured in a striped phase which breaks rotational symmetry 
(as discussed below).  In the following, we choose $\beta$ large enough to ensure
simulation of ground-state properties, and the system size is $N=L\times L$.

\begin{figure}[tb]
\begin{center}
  \includegraphics[angle=0,width=3.3in,clip]{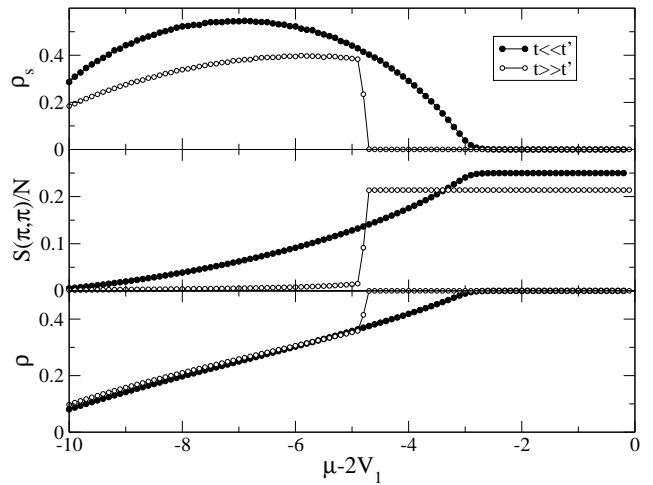}
\end{center}
\caption{
$\rho_s$, $S(\pi,\pi)$ and the number density $\rho$ 
vs. the chemical potential $\mu$ for $V_1=3.5$ and $L=16$ at an inverse temperature $\beta=10$. All quantities are in units of $t'$. 
} 
\label{fig:SSEchecker}
\end{figure}

\begin{figure}[tb]
\begin{center}
\includegraphics[angle=0,width=3.3in,clip]{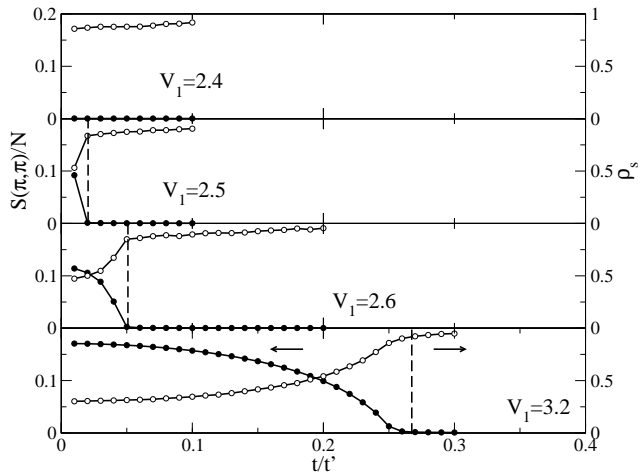}
\end{center}
\caption{
$S(\pi,\pi)$(solid symbols)  and  $\rho_s$(open symbols) vs. the nn 
hopping $t$. $\mu-2V_1=-3.5, \beta=10$ and $L=32$. 
The vertical dash line divides the two phases: the right-hand side is a superfluid state and the left-hand side is a checkerboard supersolid state. All quantities are in units of $t'$.
}
\label{fig:SSEmelt}
\end{figure}

We begin by examining the phase transition into the cSS state, identified in Fig.~\ref{fig:MF1}.
In the limit where $V_2$ vanishes, Fig.~\ref{fig:SSEchecker} shows the behavior of the QMC observables at $V_1=3.5t'$.  For $t \gg t' (t=100 
t')$ (open symbols), there is a discontinuity near 
$\mu-2V_1=-5t'$, where
a checkerboard solid with finite $S(\pi,\pi)$ melts into a superfluid with finite $\rho_s$ via a first-order transition. 
This discontinuous jump in the particle density near $\mu-2V_1=-5t'$ is a clear indication that phase separation would occur in a canonical system.\cite{Batrouni00PRL}
In contrast, in the limit $t \ll t' (t=0.01t')$ (solid symbols), the discontinuity disappears, and a smooth decrease in the structure factor as holes 
are doped into the system is accompanied by an increasing superfluidity. 
The coexistence of both finite $S(\pi,\pi)$ and $\rho_s$ in contrast to the case 
$t'=0$, indicates that a checkerboard supersolid state is stabilized by the nnn hopping.
In order to confirm that this is indeed true, we perform simulations with finite nn hopping $t$.
In Fig.~\ref{fig:SSEmelt}, we show QMC results as a function of nn hopping $t$ for various $V_1$ and the chemical potential fixed at 
$\mu-V_1=-3.5t'$. 
A supersolid phase emerges for $V_1>2.5t'$, and  a checkerboard supersolid to superfluid transition occurs as $t$ increases. 
The smooth nature of the data across the transition region suggests that the destabilization of the cSS state 
upon increased $t$ occurs via a continuous phase transition.

\begin{figure}[tb]
\begin{center}
\includegraphics[angle=0,width=3.3in,clip]{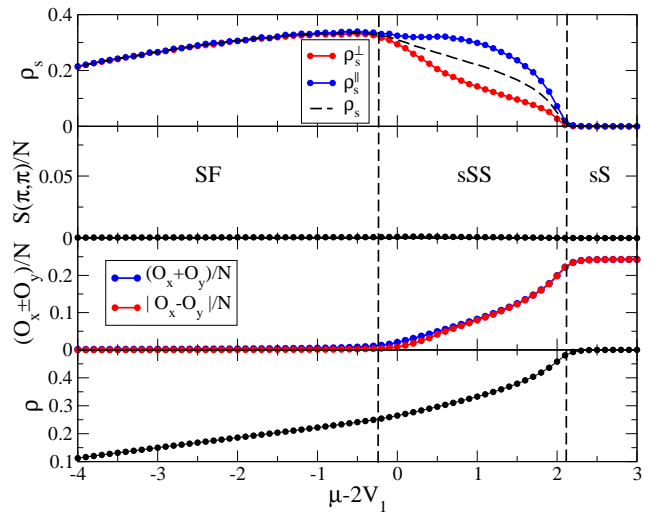}
\end{center}
\caption{(Color online). The structure factors, $\rho_s$, and $\rho$ for $x=0.9$, $V_1=1.0$,
$L=16$ and $\beta=10$. 
In the striped supersolid (sSS), $\rho_s$ shows a strong anisotropy. All quantities are in units of $t+t'$.
}
\label{fig:V10}
\end{figure}

Next, we consider the effect of the nnn repulsion $V_2$, as alluded to in Fig.~\ref{fig:MF2}.
We focus on the results from simulations performed at $x=0.9$, corresponding to $t=0.9, t'=0.1$ and $V_2=4.5$. 
Three different 
values of the nn repulsion are chosen: $V_1=1.0, 3.0$ and $4.5$. 
For $V_1=1.0$, the dominant $t$ and $V_2$ render the model close to a $t-V_2$ model, and a striped structure is expected (Fig.~\ref{fig:MF2}).
In Fig.~\ref{fig:V10}, the equivalence of $O_x+O_y$ and $|O_x-O_y|$ indicates the absence of the quarter-filled structure, and indeed,
at half-filling, a stable striped solid (sS) is formed. 
Furthermore, upon hole-doping away from $\rho=1/2$, a striped supersolid emerges.
To assess the behavior of the superflow in the sSS, we 
measured the superfluid densities perpendicular and parallel to the actual stripe 
direction.  For this purpose, $\rho^{\perp}_s$ and $\rho^{\parallel}_s$ are defined by comparing the 
magnitude of $O_x$ and $O_y$ calculated after each Monte Carlo step: when $O_x>O_y$, the 
$x$-direction winding number $W_x$ (see Eq.~(\ref{RHOs})) is counted as $W_{\parallel}$ and $W_y$ is 
counted as $W_{\perp}$, and vice versa \cite{Melko06PRB}.
Fig.~\ref{fig:V10} clearly exhibits 
a pronounced anisotropy of $\rho_s$ in the sSS phase. 
Upon further hole doping, we observe a melting of the crystal structure to a uniform superfluid (SF).
This completes the quantum melting of the sS crystal upon doping holes -- proceeding to a uniform superfluid state via an 
intermediate sSS state with coexisting superflow and crystal order.

\begin{figure}[bt]
\begin{center}
\includegraphics[angle=0,width=3.3in,clip]{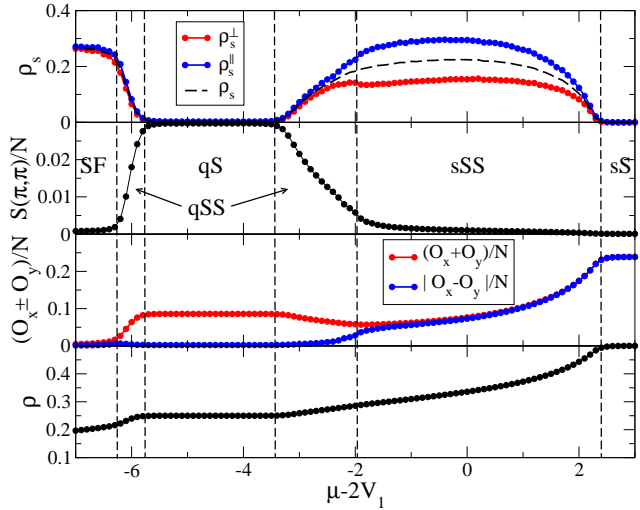}
\end{center}
\caption{
(Color online).The structure factors, $\rho_s$, and $\rho$  for $x=0.9$ and
$V_1=4.5$. The size is $L=16$ with inverse temperature $\beta=10$. Quarter-filled supersolids (qSS) and a solid (qS) are 
observed near $\rho=1/4$. 
A qSS to sSS transition is observed upon doping (see Fig.~\ref{fig:transition}).  $\rho_s$ also shows strong
anisotropies inside the sSS phase. All quantities are in units of $t+t'$.
}
\label{fig:V45}
\end{figure}

To stabilize the quarter-filled solid, we  study  the system with a strong nn repulsion.
Fig.~\ref{fig:V45} shows the results for $V_1=4.5$. A quarter-filled solid (qS) is stabilized at $\rho=1/4$, whereas at half-filling, a sS is formed. Doping away from 
quarter-filling with holes, a ``quarter-filled'' supersolid (qSS) state is formed,\cite{qSS} as signified by the coexistence of the quarter-filled crystal structure and superfluidity. 
Upon further hole-doping, the qSS eventually melts into a SF. Doping slightly away from quarter-filling with additional bosons, we observe a similar qSS state. With 
further doping, however, $O_x$ and $O_y$, as well as $\rho_s$, begin to exhibit significant anisotropies. 

\begin{figure}[bt]
\begin{center}
\includegraphics[width=3.3in]{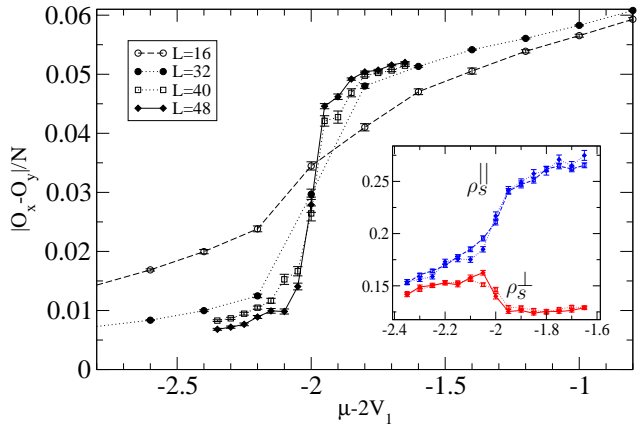}
\end{center}
\caption{ (Color online).
Detail of the structure factor difference (main) and superfluid density anisotropy (inset) for the simulations discussed in Fig.~\ref{fig:V45}, where the discontinuities developing with increased system size indicate a first-order phase transition between the ``quarter-filled'' and striped supersolids. 
}
\label{fig:transition}
\end{figure}

Near $\mu-2V_1\approx -2$, the anisotropy is most pronounced, and $S(\pi,\pi)$ vanishes, signifying a sSS state. We thus observe two seemingly unique 
supersolid states with  different underlying crystal structures.  A detailed study of the transition region
in Fig.~\ref{fig:transition} indicates the presence of discontinuities developing in the structure factors
and superfluid density at the transition.  This indicates a first-order phase transition between
the two supersolid phases, as traversed by varying the chemical potential.  In a simple phenomenological defect-condensation picture, this transition may be 
interpreted as occurring via the first-order melting of one of the crystal sublattices 
that differentiate the qSS from the sSS.  This interpretation is discussed more in the next section.


\begin{figure}[bt]
\begin{center}
\includegraphics[angle=0,width=3.3in,clip]{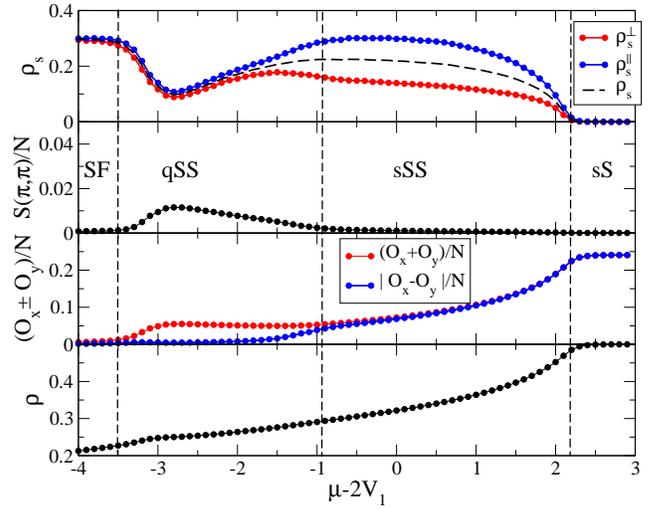}
\end{center}
\caption{(Color online).
The structure factors, $\rho_s$, and $\rho$ for $x=0.9$ and $V_1=3.0$. The
size is $L=16$ with inverse temperature is $\beta=10$. All quantities are in units of $t+t'$.
}
\label{fig:V30}
\end{figure}

\begin{figure}[bt]
\begin{center}
\includegraphics[angle=0,width=3.3in,clip]{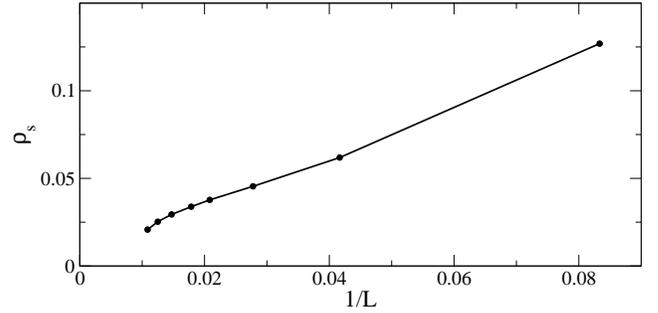}
\end{center}
\caption{
Finite size behavior of  $\rho_s$ at $\rho=1/4$ (see
Fig.~\ref{fig:V30}). The solid curve is a guide to the eye. 
}
\label{fig:FSS}
\end{figure}

In Fig.~\ref{fig:V30}, with slightly smaller $V_1=3.0$, the qSS state is still 
observed, yet with a reduced extent; no obvious qS crystal is observed at quarter-filling on this lattice size.  However, the superfluid density, although finite, shows a large dip near $\mu-2V_1=2.9$ where the average particle density nears $\rho=1/4$.  
In order to examine this more precisely, we performed simulations
at a fixed particle density $\rho=1/4$, by carefully adjusting the chemical potential, and restricting measurements to 
those Monte Carlo configurations with a particle number that precisely matches $\rho=1/4$.
The data in Fig.~\ref{fig:FSS} strongly suggests that the superfluid density indeed scales to zero in the 
thermodynamic limit, revealing the absence of supersolid behavior at $\rho=1/4$.
This observation is consistent with the picture of supersolidity in this model occurring
{\em only} away from commensurate crystal fillings, and arising due to the superflow of doped defects placed interstitial to the ordered solid structures.

\section{Discussion}

Using mean-field theory and quantum Monte Carlo simulations, we studied in detail
the formation of three supersolid phases, 
which arise in the hard-core extended boson Hubbard model of Eq.~(1).
For large nnn repulsion, a stable checkerboard supersolid phase can be observed
provided a sufficiently strong nnn hopping is present, if the system is doped away from commensurate
(1/2) filling.  As observed in previous studies, the nn hopping itself is {\it not} sufficient to promote superflow within the doped checkerboard crystal.
The other 1/2-filled crystal observed in this model is the striped solid that breaks rotational symmetry.  
Again, upon doping away from half-filling, a supersolid state emerges from the striped solid.
Furthermore, at lower density and large repulsive interactions ($V_1$ and $V_2$), the underlying density order changes to 
a quarter-filled crystal structure in order to avoid the large repulsions.  
At particle density of exactly 1/4, traces of the superflow vanish.

The above observations lend strong support to the idea of a mechanism for supersolidity involving the condensation of dopants (defects) 
outside of the lattice crystal.  Indeed, in no instance can we successfully stabilize both a 
finite crystal order parameter {\it and} a superfluid density at any commensurate filling.  
However, the simple phenomenological picture of the doped-defect condensation clearly breaks down at least for the striped supersolid phase, 
where although $\rho^{\perp}_s$ and $\rho^{\parallel}_s$ show a strong anisotropy, $\rho^{\perp}_s$ remains finite even close to  the half-filled striped crystal.  
This demonstrates that the superfluidity in the striped supersolid is {\it not} merely a one-dimensional superflow through one-dimensional channels.  This finding is similar to observations in other models on the square lattice,\cite{Hebert01PRB,SchmidPRL04, Sengupta05PRL} and contrasts to the very weak anisotropies observed on a triangular lattice striped supersolid at half-filling.\cite{Melko06PRB} 

The presence of different supersolid phases in this model also raises the interesting possibility
of  observing direct supersolid-supersolid phase transitions.  In particular, upon tuning $x$, we  studied the 
intermediate region between the $t'-V_1$ and the $t-V_2$ model.  We find that there is no direct transition 
between the checkerboard and the striped supersolid orders as $x$ is tuned -- there is always a 
superfluid phase present when the repulsions become comparable.\cite{ChenMaster} 
This is similar to the case at half-filling, where the superfluid emerges 
along the line $V_1=2V_2$ without a direct transition between the 
checkerboard and striped solid, even when both $V_1$ and $V_2$ are large.\cite{Batrouni95PRL}

In contrast, we find a direct transition between the qSS and sSS states in this model upon tuning $\mu-2V_1$.  
A detailed finite size study reveals that this supersolid-supersolid phase transition is a first-order
stripe melting transition.  Tuning 
from the sSS towards the qSS by decreasing the chemical potential, an abrupt increase in 
the superfluid density component perpendicular to the stripe direction takes place, corresponding to a jump
into the qSS crystal structure.  This observation lends itself to the interpretation that, upon
traversing this phase boundary, one of the two occupied sublattices, that contribute to 
the striped crystal, abruptly melts into a superfluid component, while the other remains its rigidity,
and provides the underlying qSS crystal structure.  It would be interesting to compare this mechanism
to that observed in a supersolid-supersolid phase transition on the 
triangular lattice,\cite{Tizzle} where significantly stronger 
first-order behavior is observed.  There have also been proposed more exotic mechanisms, where superfluids transition into non-uniform solid phases at commensurate filling, which may be compared to the current work.\cite{DVT}

In conclusion, we have found several ground state phases of the hard-core extended boson Hubbard model with nn and nnn hopping and repulsion on the square lattice.
Most notable, we find that supersolid states readily emerge when doped away from commensurability ``near'' their associated crystal phases, provided sufficient kinetic (hopping) freedom is provided.  
The model thus proves an ideal playground for future study of concepts related to doping and the formation of supersolidity through the mechanism of condensed defects.
Further studies are necessary to understand the detailed nature of the transitions between these different solid, superfluid and supersolid phases, 
as well as their finite temperature properties.

\begin{acknowledgments}
This work was supported by NSC and NCTS of Taiwan (YCC,YJK),
the U.S. Department of Energy, contract DE-AC05-00OR22725 with Oak Ridge National Laboratory,
managed by UT-Battelle, LLC (RGM), the German
Research Foundation, NIC J\"ulich and HLRS Stuttgart (SW)
and the National Science Foundation
under Grand No. NSF PHYS05-51164 (SW,YJK).
RGM would like to thank the Center of Theoretical Sciences and Department of Physics, National Taiwan University, for 
the hospitality extended during a visit, and SW and YJK 
acknowledge hospitality of the Kavli Institute for Theoretical Physics at Santa Barbara.
\end{acknowledgments}

\bibliographystyle{apsrev}
\bibliography{./refSS}

\begin{thebibliography}{31}
\expandafter\ifx\csname natexlab\endcsname\relax\def\natexlab#1{#1}\fi
\expandafter\ifx\csname bibnamefont\endcsname\relax
  \def\bibnamefont#1{#1}\fi
\expandafter\ifx\csname bibfnamefont\endcsname\relax
  \def\bibfnamefont#1{#1}\fi
\expandafter\ifx\csname citenamefont\endcsname\relax
  \def\citenamefont#1{#1}\fi
\expandafter\ifx\csname url\endcsname\relax
  \def\url#1{\texttt{#1}}\fi
\expandafter\ifx\csname urlprefix\endcsname\relax\def\urlprefix{URL }\fi
\providecommand{\bibinfo}[2]{#2}
\providecommand{\eprint}[2][]{\url{#2}}

\bibitem[{\citenamefont{Penrose and Onsager}(1956)}]{Penrose56PR}
\bibinfo{author}{\bibfnamefont{O.}~\bibnamefont{Penrose}} \bibnamefont{and}
  \bibinfo{author}{\bibfnamefont{L.}~\bibnamefont{Onsager}},
  \bibinfo{journal}{Phys. Rev.} \textbf{\bibinfo{volume}{104}},
  \bibinfo{pages}{576} (\bibinfo{year}{1956}).

\bibitem[{\citenamefont{Andreev and Lifshits}(1969)}]{Andreev69JETP}
\bibinfo{author}{\bibfnamefont{A.}~\bibnamefont{Andreev}} \bibnamefont{and}
  \bibinfo{author}{\bibfnamefont{I.}~\bibnamefont{Lifshits}},
  \bibinfo{journal}{Sov. Phys. JETP} \textbf{\bibinfo{volume}{29}},
  \bibinfo{pages}{1107} (\bibinfo{year}{1969}).

\bibitem[{\citenamefont{Chester}(1970)}]{Chester70PRA}
\bibinfo{author}{\bibfnamefont{G.~V.} \bibnamefont{Chester}},
  \bibinfo{journal}{Phys. Rev. A} \textbf{\bibinfo{volume}{2}},
  \bibinfo{pages}{256} (\bibinfo{year}{1970}).

\bibitem[{\citenamefont{Leggett}(1970)}]{Leggett70PRL}
\bibinfo{author}{\bibfnamefont{A.~J.} \bibnamefont{Leggett}},
  \bibinfo{journal}{Phys. Rev. Lett.} \textbf{\bibinfo{volume}{25}},
  \bibinfo{pages}{1543} (\bibinfo{year}{1970}).

\bibitem[{\citenamefont{Kim and M.H.W.Chan}(2004)}]{Kim04Nature}
\bibinfo{author}{\bibfnamefont{E.}~\bibnamefont{Kim}} \bibnamefont{and}
  \bibinfo{author}{\bibnamefont{M.H.W.Chan}}, \bibinfo{journal}{Nature}
  \textbf{\bibinfo{volume}{427}}, \bibinfo{pages}{225} (\bibinfo{year}{2004}).

\bibitem[{\citenamefont{Ceperley and Bernu}(2004)}]{Ceperley04PRL}
\bibinfo{author}{\bibfnamefont{D.~M.} \bibnamefont{Ceperley}} \bibnamefont{and}
  \bibinfo{author}{\bibfnamefont{B.}~\bibnamefont{Bernu}},
  \bibinfo{journal}{\prl} \textbf{\bibinfo{volume}{93}}, \bibinfo{eid}{155303}
  (\bibinfo{year}{2004}).

\bibitem[{\citenamefont{Prokof'ev and Svistunov}(2005)}]{Prokofev05PRL}
\bibinfo{author}{\bibfnamefont{N.}~\bibnamefont{Prokof'ev}} \bibnamefont{and}
  \bibinfo{author}{\bibfnamefont{B.}~\bibnamefont{Svistunov}},
  \bibinfo{journal}{\prl} \textbf{\bibinfo{volume}{94}}, \bibinfo{eid}{155302}
  (\bibinfo{year}{2005}).

\bibitem[{\citenamefont{Burovski et~al.}(2005)}]{Burovski05PRL}
\bibinfo{author}{\bibfnamefont{E.}~\bibnamefont{Burovski}}
  \bibnamefont{et~al.}, \bibinfo{journal}{\prl} \textbf{\bibinfo{volume}{94}},
  \bibinfo{eid}{165301} (\bibinfo{year}{2005}).

\bibitem[{\citenamefont{Kim and Chan}(2006)}]{Kim06PRL}
\bibinfo{author}{\bibfnamefont{E.}~\bibnamefont{Kim}} \bibnamefont{and}
  \bibinfo{author}{\bibfnamefont{M.~H.~W.} \bibnamefont{Chan}},
  \bibinfo{journal}{\prl} \textbf{\bibinfo{volume}{97}}, \bibinfo{eid}{115302}
  (\bibinfo{year}{2006}).

\bibitem[{\citenamefont{Rittner and Reppy}(2006)}]{Rittner06arxiv}
\bibinfo{author}{\bibfnamefont{A.~S.~C.} \bibnamefont{Rittner}}
  \bibnamefont{and} \bibinfo{author}{\bibfnamefont{J.~D.} \bibnamefont{Reppy}},
  \bibinfo{journal}{\prl} \textbf{\bibinfo{volume}{97}},
  \bibinfo{pages}{165301} (\bibinfo{year}{2006}).

\bibitem[{\citenamefont{Day and Beamish}(2006)}]{day:105304}
\bibinfo{author}{\bibfnamefont{J.}~\bibnamefont{Day}} \bibnamefont{and}
  \bibinfo{author}{\bibfnamefont{J.}~\bibnamefont{Beamish}},
  \bibinfo{journal}{\prl} \textbf{\bibinfo{volume}{96}},
  \bibinfo{pages}{105304} (\bibinfo{year}{2006}).

\bibitem[{\citenamefont{Anderson et~al.}(1995)}]{Anderson95Science}
\bibinfo{author}{\bibfnamefont{M.~H.} \bibnamefont{Anderson}}
  \bibnamefont{et~al.}, \bibinfo{journal}{Science}
  \textbf{\bibinfo{volume}{269}}, \bibinfo{pages}{198} (\bibinfo{year}{1995}).

\bibitem[{\citenamefont{Greiner et~al.}(2002)}]{Greiner02Nature}
\bibinfo{author}{\bibfnamefont{M.}~\bibnamefont{Greiner}} \bibnamefont{et~al.},
  \bibinfo{journal}{Nature} \textbf{\bibinfo{volume}{415}}, \bibinfo{pages}{39}
  (\bibinfo{year}{2002}).

\bibitem[{\citenamefont{Griesmaier et~al.}(2005)}]{Griesmaier05PRL}
\bibinfo{author}{\bibfnamefont{A.}~\bibnamefont{Griesmaier}}
  \bibnamefont{et~al.}, \bibinfo{journal}{\prl} \textbf{\bibinfo{volume}{94}},
  \bibinfo{eid}{160401} (\bibinfo{year}{2005}).

\bibitem[{dip()}]{dipolar_theory}
\bibinfo{note}{D.~Jaksch \textit{et al.}, Phys. Rev. Lett. \textbf{81}, 3108
  (1998); K.~G\'oral, L.~Santos and M.~Lewenstein, \textit{ibid.} \textbf{88},
  170406 (2002); V.~W. Scarola and S.~Das~Sarma, \textit{ibid.}, \textbf{95}
  033003 (2005)}.

\bibitem[{\citenamefont{Batrouni et~al.}(1995)}]{Batrouni95PRL}
\bibinfo{author}{\bibfnamefont{G.~G.} \bibnamefont{Batrouni}}
  \bibnamefont{et~al.}, \bibinfo{journal}{Phys. Rev. Lett.}
  \textbf{\bibinfo{volume}{74}}, \bibinfo{pages}{2527} (\bibinfo{year}{1995}).

\bibitem[{\citenamefont{Sengupta et~al.}(2005)}]{Sengupta05PRL}
\bibinfo{author}{\bibfnamefont{P.}~\bibnamefont{Sengupta}}
  \bibnamefont{et~al.}, \bibinfo{journal}{\prl} \textbf{\bibinfo{volume}{94}},
  \bibinfo{eid}{207202} (\bibinfo{year}{2005}).

\bibitem[{\citenamefont{H\'ebert et~al.}(2001)}]{Hebert01PRB}
\bibinfo{author}{\bibfnamefont{F.}~\bibnamefont{H\'ebert}}
  \bibnamefont{et~al.}, \bibinfo{journal}{Phys. Rev. B}
  \textbf{\bibinfo{volume}{65}}, \bibinfo{pages}{014513}
  (\bibinfo{year}{2001}).

\bibitem[{tri()}]{triangular_SS}
\bibinfo{note}{S.~Wessel and M.~Troyer, Phys. Rev. Lett. \textbf{95},
  127205(2005);D.~Heidarian and K.~Damle, \textit{ibid.} \textbf{95}, 127206
  (2005);R. G. Melko, et al., \textit{ibid.},\textbf{95}, 127207 (2005);
  M.~Boninsegni and N. Prokof'ev, \textit{ibid.}, \textbf{95}, 238204 (2005)}.

\bibitem[{kag()}]{kagome_SS}
\bibinfo{note}{S.~Wessel, Phys. Rev. B \textbf{75}, 174301(2007);J.-Y. Gan
  \textit{et al.},\textit{ibid.} \textbf{75},214509 (2007)}.

\bibitem[{\citenamefont{Melko et~al.}(2006)\citenamefont{Melko, {Del Maestro},
  and Burkov}}]{Melko06PRB}
\bibinfo{author}{\bibfnamefont{R.~G.} \bibnamefont{Melko}},
  \bibinfo{author}{\bibfnamefont{A.}~\bibnamefont{{Del Maestro}}},
  \bibnamefont{and} \bibinfo{author}{\bibfnamefont{A.~A.}
  \bibnamefont{Burkov}}, \bibinfo{journal}{\prb} \textbf{\bibinfo{volume}{74}},
  \bibinfo{pages}{214517} (\bibinfo{year}{2006}).

\bibitem[{\citenamefont{Hassan et~al.}(2007)\citenamefont{Hassan, de~Medici,
  and Tremblay}}]{Tizzle}
\bibinfo{author}{\bibfnamefont{S.~R.} \bibnamefont{Hassan}},
  \bibinfo{author}{\bibfnamefont{L.}~\bibnamefont{de~Medici}},
  \bibnamefont{and} \bibinfo{author}{\bibfnamefont{A.-M.}
  \bibnamefont{Tremblay}} (\bibinfo{year}{2007}), \eprint{arXiv:0707.0866v1}.

\bibitem[{\citenamefont{Batrouni and Scalettar}(2000)}]{Batrouni00PRL}
\bibinfo{author}{\bibfnamefont{G.~G.} \bibnamefont{Batrouni}} \bibnamefont{and}
  \bibinfo{author}{\bibfnamefont{R.~T.} \bibnamefont{Scalettar}},
  \bibinfo{journal}{Phys. Rev. Lett.} \textbf{\bibinfo{volume}{84}},
  \bibinfo{pages}{1599} (\bibinfo{year}{2000}).

\bibitem[{\citenamefont{Sandvik and Kurkij\"arvi}(1991)}]{Sandvik91PRB}
\bibinfo{author}{\bibfnamefont{A.~W.} \bibnamefont{Sandvik}} \bibnamefont{and}
  \bibinfo{author}{\bibfnamefont{J.}~\bibnamefont{Kurkij\"arvi}},
  \bibinfo{journal}{Phys. Rev. B} \textbf{\bibinfo{volume}{43}},
  \bibinfo{pages}{5950} (\bibinfo{year}{1991}).

\bibitem[{\citenamefont{Syljuasen and Sandvik}(2002)}]{Syljuasen02PRE}
\bibinfo{author}{\bibfnamefont{O.~F.} \bibnamefont{Syljuasen}}
  \bibnamefont{and} \bibinfo{author}{\bibfnamefont{A.~W.}
  \bibnamefont{Sandvik}}, \bibinfo{journal}{Phys. Rev. E}
  \textbf{\bibinfo{volume}{66}}, \bibinfo{pages}{046701}
  (\bibinfo{year}{2002}).

\bibitem[{\citenamefont{Schmid}(2004)}]{Schmid04phd}
\bibinfo{author}{\bibfnamefont{G.}~\bibnamefont{Schmid}}, Ph.D. thesis,
  \bibinfo{school}{ETH, Z\"{u}rich} (\bibinfo{year}{2004}).

\bibitem[{\citenamefont{Gutzwiller}(1963)}]{Gutzwiller63PRL}
\bibinfo{author}{\bibfnamefont{M.~C.} \bibnamefont{Gutzwiller}},
  \bibinfo{journal}{Phys. Rev. Lett.} \textbf{\bibinfo{volume}{10}},
  \bibinfo{pages}{159} (\bibinfo{year}{1963}).

\bibitem[{\citenamefont{Schmid and Troyer}(2004)}]{SchmidPRL04}
\bibinfo{author}{\bibfnamefont{G.}~\bibnamefont{Schmid}} \bibnamefont{and}
  \bibinfo{author}{\bibfnamefont{M.}~\bibnamefont{Troyer}},
  \bibinfo{journal}{Phys. Rev. Lett.} \textbf{\bibinfo{volume}{93}}
  (\bibinfo{year}{2004}).

\bibitem[{qSS()}]{qSS}
\bibinfo{note}{Note, that the qSS is close to but not not exactly at
  $\rho=1/4$.}

\bibitem[{Che()}]{ChenMaster}
\bibinfo{note}{Y.~C. Chen, Master thesis, National Taiwan University, Taipei,
  (2007)}.

\bibitem[{DVT()}]{DVT}
\bibinfo{note}{L.~Balents, L.~Bartosch, A.~Burkov, S.~Sachdev and K.~Sengupta,
  Phys. Rev. B \textbf{71} 144508; \textit{ibid.} \textbf{71} 144509 (2005)}.

\end{thebibliography}

\end{document}